\documentclass[runningheads]{llncs}

\usepackage[T1]{fontenc}
\usepackage[utf8]{inputenc}
\usepackage[english]{babel}
\usepackage{amsmath,amssymb}
\usepackage{mathtools}
\usepackage{bbm}
\usepackage{graphicx}
\usepackage{booktabs}
\usepackage{xcolor}
\usepackage{enumitem}
\usepackage{microtype}
\usepackage[hidelinks]{hyperref}

\newcommand{\PP}{\mathbb{P}}
\newcommand{\EE}{\mathbb{E}}
\newcommand{\RR}{\mathbb{R}}
\newcommand{\1}{\mathbbm{1}}
\newcommand{\cL}{\mathcal{L}}
\newcommand{\cX}{\mathcal{X}}
\newcommand{\hatt}{\hat{\tau}}
\newcommand{\hmu}{\hat{\mu}}
\newcommand{\he}{\hat{e}}
\newcommand{\sign}{\operatorname{sign}}

\begin{document}

\title{Should They Shoot?}
\subtitle{Behavioral regularization for decision-focused causal learning in football under low overlap}
\titlerunning{Should They Shoot? Behavioral-regularized causal learning}

%
\author{Rémi Verronneau\inst{1,2} \and
Matías Ortiz-Angel\inst{3} \and
Vianney Perchet\inst{2}}
\authorrunning{R. Verronneau et al.}
\institute{
Footovision \email{remi.verronneau@footovision.com}
\and
CREST, ENSAE, \& Inria FairPlay
\and
Universidad de Chile
}
\maketitle              
%

\begin{abstract}
We study the shooting decision in football from observational data. The goal is not only to predict a well-calibrated goal probability, but to learn a causal decision rule for \emph{when} a player should shoot. Here, policy learning based on the Conditional Average Treatment Effect (CATE) is unstable: shooting is beneficial on average, yet rarely observed in many states, especially where the propensity to shoot is low. We propose a decision-focused behavioral regularization term that uses observed expert actions as a directional signal to stabilize the sign of the treatment-effect score in low-overlap regions, the exact regime where standard causal estimators fail and where the policy decision matters most. We give a population-level analysis of the term (its minimizer is proportional to the log-odds of the behavior policy), a controlled evaluation on simulated data where the true CATE is known, and an application on recent real data using spatio-temporal graphs encoded by a graph neural network. Empirically, behavioral regularization is the only variant that achieves both factual quality comparable to DragonNet \emph{and} a realistic, actionable policy (shot rate and agreement with expert behavior close to human decisions), where the others achieve only one.
\keywords{causal inference \and policy learning \and low overlap \and graph neural networks \and football analytics.}
\end{abstract}

\section{Introduction}
\label{sec:intro}
\emph{Why doesn't he shoot?} In professional football, deciding whether to shoot or keep the ball is a repeated, high-stakes decision, especially in the dangerous areas of the pitch. These decisions directly affect match outcomes and sit at the heart of tactical analysis. Existing tools such as \emph{expected goals} (xG) or models that assign value to actions and possessions (VAEP, EPV)~\cite{decroos2019actions,fernandez2021framework} give useful descriptions, but they do not answer a causal question: {given the positions, velocities, and defensive context of the current frame, would the team be better off if the player shot? In short: \emph{given the current game state, should the player shoot?}}

This question is naturally framed in terms of conditional treatment effects. Let $T=1$ be the action \emph{shoot} and $T=0$ \emph{do not shoot}. For a game state $x$, the conditional average treatment effect (CATE) is $\tau(x)=\EE[Y(1)-Y(0)\mid X=x]$, where $Y(1),Y(0)$ are the potential outcomes (a goal within a fixed horizon), and the matching decision rule is $\hat\pi(x)=\1\{\hatt(x)>0\}$. A problem shows up right away: although the average effect of shooting is often positive, shooting is \emph{conditionally} rare in many states. This low-overlap regime~\cite{crump2009dealing} makes local comparisons between actions unreliable. A standard CATE estimator can then assign positive scores almost everywhere and produce an unrealistic ``always shoot'' policy, which is exactly the failure mode we want to avoid in decision support.

Our proposal is the following: we augment a standard causal objective (DragonNet)~\cite{shi2019dragonnet} with a \emph{behavioral regularization} term built from the observed action. The regularizer does not replace the causal objective: it adds a directional learning signal. {In professional football, players do not shoot at random: on average, they shoot more often in states where shooting is beneficial. This means that the observed action carries weak but useful information about the sign of the treatment effect, which we exploit as a directional signal to stabilize the policy in low-overlap regions.}

\paragraph{Contributions.}
\textbf{(C1)} We frame the shooting decision as a low-overlap policy-learning problem from observational data. \textbf{(C2)} We propose a behavioral-regularized DragonNet on game graphs that uses the observed action as a directional regularizer of the CATE score. \textbf{(C3)} We characterize, at the population level, the minimizer of the behavioral term, which is proportional to the log-odds of the behavior policy. \textbf{(C4)} We validate the mechanism on simulated data (true CATE known) and then on real multi-league data, showing that behavioral regularization gives the best trade-off between factual quality and decision quality.

\section{Setup and problem}
\label{sec:problem}
We observe i.i.d.\ triples $(X_i,T_i,Y_i)_{i=1}^n$, with $X\in\cX$ the game state, $T\in\{0,1\}$ the observed action, and $Y\in\{0,1\}$ the outcome. Let $Y(1),Y(0)$ be the potential outcomes and $e(x):=\PP(T=1\mid X=x)$ the propensity. We assume \emph{consistency} ($Y=TY(1)+(1-T)Y(0)$) and \emph{{positivity}} {($0 < e(x) < 1$ for all $x\in\cX$). In practice, however, shooting is a conditionally rare action: far from goal or under heavy defensive pressure, players almost never shoot ($e(x)\approx 0$), while in clear one-on-one situations they almost always do ($e(x)\approx 1$). This is precisely the low-overlap regime motivated in Section~\ref{sec:intro} and the central challenge this work addresses.}

{Unlike most causal inference methods, we do \emph{not} assume \emph{conditional ignorability} ($Y(0),Y(1)\perp T\mid X$).} In practice, unobserved factors (field vision, instantaneous perceptual information, communication between teammates, or the player's intent) can affect both the decision to shoot and the chance of success. We therefore separate three levels. \textbf{(i)} The \emph{ideal causal quantity} $\tau(x)=\EE[Y(1)-Y(0)\mid X=x]$ defines the optimal policy $\pi^\star(x)=\1\{\tau(x)>0\}$; we do not assume it is identifiable. \textbf{(ii)} The \emph{estimated observational quantity} $\tilde\tau(x){:=\mu_1(x)-\mu_0(x)}${, where $\mu_t(x):=\EE[Y\mid T{=}t,X{=}x]$ are the conditional outcome means,} is what the network actually learns{, producing an estimate $\hat{\tilde\tau}(x)$. In general, $\tilde\tau(x)=\tau(x)+B(x)$, where $B(x)$ is the selection bias concentrated in the low-overlap region (characterized empirically in Sec.~\ref{sec:sim}); note that under conditional ignorability $B(x)=0$ and $\tilde\tau=\tau$, but we do not assume this holds here.} \textbf{(iii)} {The ultimate target is $\sign(\tau(x))$; since $\tau$ is not identifiable, the} \emph{real objective} is the \emph{sign} of $\hat{\tilde\tau}${, which agrees with $\sign(\tau(x))$ whenever $B(x)$ does not flip it:} behavioral regularization aims to fix $\sign(\hat{\tilde\tau})$ where $B(x)$ flips it, without claiming to recover $\tau(x)$. {The lack of conditional ignorability is not a limitation of our approach: our goal is not to identify $\tau(x)$ exactly, but to learn a decision rule with the correct sign.}

\begin{figure}[h]
\centering
\includegraphics[width=0.5\textwidth]{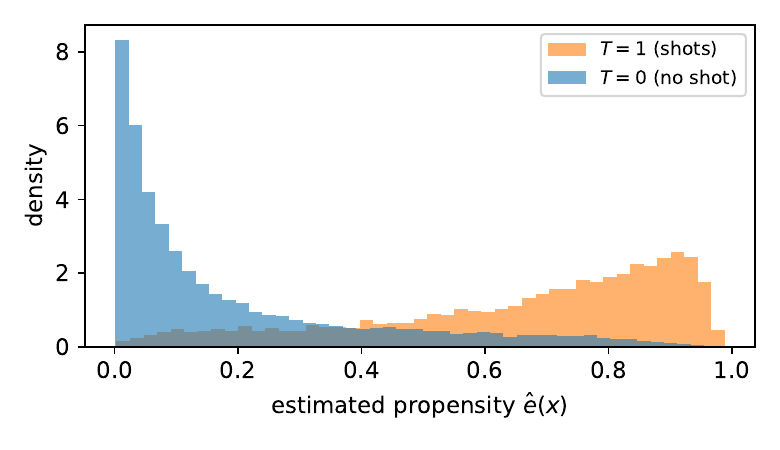}
\caption{Low overlap (real data): distribution of $\he(x)$ by action. A large mass near $0$/$1$ documents the hard regime, where causal supervision for one arm is rare.}
\label{fig:overlap}
\end{figure}

The practical difficulty is not the marginal shot rate but \emph{conditional} rarity: there are large regions where $e(x)\approx 0$. Estimating $\mu_1(x)$ there is hard, because few comparable shots are observed. This is exactly where a policy based on a naive CATE becomes unstable; the proposed regularization is a methodological fix for this regime. On our real data, $25\%$ of states have $\he(x)\notin[0.05,0.95]$ (median $0.21$, $5$th percentile $0.01$): the low-overlap regime is documented empirically (Fig.~\ref{fig:overlap}).

\section{Related work}
\label{sec:related}
\paragraph{Causal effect estimation and policy learning.} The potential-outcomes framework goes back to Rubin~\cite{rubin1974} and Rosenbaum--Rubin~\cite{rosenbaum1983}, who introduced the propensity score as the key balancing quantity in observational studies. A large family of CATE estimators followed: meta-learners that split the problem into nuisance regressions (S-, T-, and X-learners) and tree-based methods such as causal forests, alongside neural estimators like TARNet and DragonNet that learn a shared representation with outcome and propensity heads and add a targeted correction~\cite{shalit2017,shi2019dragonnet}. Policy learning from observational data~\cite{athey2021policy,zhao2012owl} then turns such an estimate into a decision rule. All of these methods aim at recovering $\tau(x)$ accurately, which presupposes identifiability; when an unobserved confounder is present, they all learn the same biased $\tilde\tau=\tau+B$ and offer no mechanism to correct its sign. They are therefore not the right comparison for our question, namely decision stability under hidden confounding, but variants of the naive estimator we set out to fix.
\paragraph{Low overlap and decision-making.} Limited overlap is a classic difficulty~\cite{crump2009dealing}: in these regions one treatment arm is rarely observed given $X$, inverse-propensity weights blow up, and outcome models extrapolate. Balancing methods such as MMD-based representation learning or matching rely on sufficient common support between treated and control groups~\cite{gretton2012mmd,johansson2016counterfactual,abadie2006matching,stuart2010matching,ho2007matching}. Under weak overlap, they can only extrapolate or discard the low-propensity states that are central to our decision problem.
A recent line of work argues that CATE accuracy and decision quality are distinct objectives: a policy only depends on the sign of the effect near the decision boundary, whereas standard CATE losses also spend capacity on regions where additional precision does not change the action~\cite{frauen2025,policy_aligned2025}.
We follow this view and target decision stability rather than exact recovery of $\tau(x)$.

{\paragraph{Decision support in football.} Recent football analytics has moved from valuing observed actions toward recommending alternative decisions. Rahimian et al.~\cite{rahimian2022beyond} use tracking data and deep reinforcement learning to compare a team's observed behavior with an optimized policy over ball destinations. Their question is close in spirit to ours: given a game state, which action would have improved the team's expected outcome? Van Roy et al.~\cite{vanroy2023markov} propose a broader Markov framework for learning and reasoning about team strategies from event data; one of their use cases analyzes counterfactual changes to long-distance shot policies, including whether to shoot immediately or move the ball to seek a better chance. Our work shares this decision-support motivation, but differs in its causal formulation: we isolate the shoot/don't-shoot decision and study policy learning under low overlap and hidden
confounding.}

\paragraph{Behavior regularization and football analytics.} In offline reinforcement learning, the learned policy is often pulled toward the behavior policy to curb extrapolation error~\cite{fujimoto2019,kumar2020cql}; our regularizer shares this intuition but applies it to the \emph{sign} of a treatment-effect score rather than to action values. On the football side, xG and possession-value models~\cite{anzer2021goal,fernandez2019decomposing,decroos2019actions,fernandez2021framework} are widely used to describe and value play, but they are predictive rather than causal estimators of the shooting decision. We encode game states as graphs with a graph neural network~\cite{xu2019gin,hu2020gine,fey2019pyg}.

\section{Method}
\label{sec:method}
\subsection{Graph representation of the game state}
\label{sec:graph}
Each decision is a game frame with 22 players and the ball, represented as a graph $G=(V,E)$ whose nodes are the players. Each player carries $23$ attributes (position, velocity, acceleration, distance and angle to goal, distance to the nearest defender, defenders in the shooting cone, team and role flags, ball-carrier and shooting-cone indicators), with $\sin/\cos$ encoding for angles. Edges encode three relations (within-team, carrier-to-everyone, and the $k=3$ nearest opponents), each with a normalized distance, a direction, and a type. To capture dynamics, an observation is a sequence of three consecutive frames $[g_{t-2},g_{t-1},g_{t}]${, where $t$ indexes the decision frame and $g_{t-1}, g_{t-2}$ are the two preceding frames.} A graph encoder is a natural fit: players have no canonical ordering (the representation must be permutation-invariant) and what drives a shot (defender pressure, support, lanes to goal) is relational rather than a fixed-length vector.

The shared encoder $\Phi_\theta$ applies two GINEConv layers~\cite{hu2020gine,xu2019gin} with sum aggregation, which preserves neighborhood counts (nearby opponents, defenders in the cone). A four-part readout (ball carrier, attacker mean, defender mean, global mean) and a temporal concatenation give a permutation-invariant representation $\Phi_\theta(x)\in\RR^{d}$. Three MLP heads ($h_0,h_1$ for outcomes, $g$ for propensity) then compute, with $\sigma$ the sigmoid,
\begin{align*}
  \hmu_t(x) &= \sigma\!\big(h_{t}(\Phi_\theta(x))\big), \quad t\in\{0,1\};\\
  \he(x)    &= \sigma\!\big(g(\Phi_\theta(x))\big);\\
  \hatt(x)  &= \hmu_1(x)-\hmu_0(x).
\end{align*}
The action $T$ is never an input to $\Phi_\theta$, only a label.

DragonNet~\cite{shi2019dragonnet} adds a targeted regularization inspired by
TMLE~\cite{vanderlaan2006tmle}. Under standard causal assumptions, this update
is motivated by doubly robust estimation. Since we do not assume ignorability,
we use it here as a regularized learning objective rather than as a guarantee
of consistent causal estimation.

\subsection{Behavioral regularization}
Let the observed sign variable be $S:=2T-1\in\{-1,+1\}$ ($+1$ if shot, $-1$ otherwise), and $\phi_\beta(u):=\log(1+e^{-\beta u})$ for a temperature $\beta>0$. The DragonNet base loss combines the factual outcome, the propensity, and the targeted regularization:
\begin{multline*}
  \cL_{\mathrm{base}}(\theta) := \EE\!\Big[\big(Y-\hmu_T(X)\big)^2
  + \lambda_{\mathrm{prop}}\,\mathrm{BCE}\big(T,\he(X)\big)
  + \lambda_{\mathrm{treg}}\,\big(Y-\hmu_T^{\epsilon}(X)\big)^2\Big].
\end{multline*}
Here $\mathrm{BCE}$ denotes the binary cross-entropy of the propensity head and the last term is the targeted (TMLE) regularization on the perturbed outcome $\hmu_T^{\epsilon}$ {~\cite{vanderlaan2006tmle}}. The behavioral term and the full objective are
\begin{equation}\label{eq:behav}
\cL_{\mathrm{behav}}(\theta):=\EE\!\big[\phi_\beta\!\big(S\,\hatt^\theta(X)\big)\big],
\qquad
\cL_\lambda(\theta):=\cL_{\mathrm{base}}(\theta)+\lambda\,\cL_{\mathrm{behav}}(\theta),\ \ \lambda\ge 0.
\end{equation}
The loss penalizes disagreements between the observed action and the sign of the estimated CATE; it vanishes when $\sign(S)=\sign(\hatt)$. At the sample level, it uses only $T$, never the true $e(x)$.

{To understand what the behavioral term optimizes at the population level, note that for a fixed score $v$ at state $x$, the expected loss is $\ell_{\mathrm{behav}}(v\mid x) = e(x)\phi_\beta(v)+(1-e(x))\phi_\beta(-v)$, which penalizes $v>0$ with weight $1-e(x)$ and $v<0$ with weight $e(x)$. This function is strictly convex and has a unique minimizer $v^\star(x)=\frac{1}{\beta}\mathrm{logit}(e(x))$, whose sign agrees with $e(x)-\frac{1}{2}$ (see Appendix~\ref{app}). This is precisely the correction the model needs: in low-overlap regions where $e(x)\approx 0$, the naive causal estimator tends to assign $\hat{\tilde\tau}(x)>0$ due to the selection bias $B(x)>0$, recommending shooting almost everywhere. The minimizer $v^\star(x)$ is strongly negative there, so the regularizer pushes $\hat{\tilde\tau}(x)$ toward negative values, counteracting the bias. Conversely, where $e(x)>1/2$ it reinforces shooting. The correction is achieved using only the observed actions $T$, without requiring knowledge of $\tau(x)$ or $e(x)$.}
On its own, the regularizer pushes the score toward the sign of $e(x)-\tfrac12$: the threshold $\tfrac12$ is not imposed at the sample level, it emerges from the population average over the random action $T\mid X$. Under low overlap, where causal supervision is weak or noisy, it injects a directional signal coming from behavior: $\lambda{=}0$ gives the unregularized causal model, too large a $\lambda$ drifts toward imitating behavior, and an intermediate $\lambda$ stabilizes the policy while keeping the causal supervision. This is a deliberate bias--variance trade-off, to be selected empirically in a decision-focused way.

\section{Controlled validation on simulated data}
\label{sec:sim}
On real data, $\tau(x)$ is unknown. We therefore build a data-generating
process (DGP) where the true CATE is known and which reproduces the
confounding structure of football: an unobserved variable $U$ raises both the
propensity to shoot and the probability of scoring, so that $B(x)>0$ in
low-overlap regions.

The covariates $X\in\RR^5$ are drawn from a mixture
$0.75\,\mathcal{N}(\mathbf{0},I_5)+0.25\,\mathcal{N}(\mu_b,I_5)$ with
$\mu_b=(2,1.5,1,0.5,0)^\top$ ($n{=}8000$). The confounder
$U\sim\mathrm{Exp}(1)-1$ (zero mean, right-skewed) is independent of~$X$ and
unobserved at estimation time. The propensity is
$T\mid X{=}x,U{=}u\;\sim\;\mathrm{Bern}(\sigma(f(x)/\alpha+\gamma u))$,
$\gamma{=}1$, with
\begin{equation}\label{eq:fx}
  f(x)=-1.8+0.3\,x_1+0.2\,x_2-0.15\,x_3
       +0.5\,x_1 x_2+0.3\,|x_3{-}x_4|-0.4\,(x_2{-}1)_+\,,
\end{equation}
and the overlap parameter $\alpha\in\{0.10,\dots,1.00\}$ rescales $f$:
smaller $\alpha$ concentrates $e(x,u)$ near $0$ and~$1$. The potential
outcomes are $Y(t)\sim\mathrm{Bern}(\sigma(\ell_t+\varepsilon))$,
$\varepsilon\sim\mathcal{N}(0,1.5)$, with $\ell_0{=}y_0(x)$,
$\ell_1{=}y_0(x)+\tau_X(x)+\delta U$, $\delta{=}2.5$, and
\[
  y_0(x)=0.4\,x_1+0.25\sin(2x_2)+0.15\,x_4-3.9,\qquad
  \tau_X(x)=2\tanh(x_1{-}0.3)+x_2-4.5.
\]
The confounder enters $Y(1)$ but not $Y(0)$: a high~$U$ simultaneously makes
shooting more likely (via the propensity) and more successful (via~$\ell_1$).
The true effect $\tau(x)=\EE_U[\sigma(y_0+\tau_X+\delta
U)]-\sigma(y_0)$ is computed by Monte Carlo ($10^4$ draws); the expectation
cannot be taken inside~$\sigma$ (Jensen's inequality). In low-overlap regions
($e{<}0.1$), the selection bias is $B(x)\approx+1.94$ while
$\tau(x)\approx-0.01$: the observational difference $\tilde\tau=\tau+B$ is
positive where the true effect is negative, which is the exact source of the
'always' shoot failure. The DGP is designed so that the proposed
regularization is tested in this regime: when $B(x){>}0$ and overlap is low,
it should correct the sign of $\hat{\tilde\tau}$, and this is what the
simulation verifies.

\paragraph{Mechanism.} In a low-$e(x)$ region, observing $T{=}1$ requires a high $U$, so $\EE[U\mid T{=}1,X]>0$; the term $\delta U$ then positively biases the observed $\mu_1$. Outcome models learn $\tilde\tau(x)=\tau(x)+B(x)$ with $B(x)>0$ concentrated at low $e$: the apparent CATE turns \emph{positive} where the true one is \emph{negative}, which is the exact cause of the ``always shoot'' policy. The gradient of~\eqref{eq:behav} on $\hatt$ is $-\beta s\,\sigma(-\beta s\hatt)$: at low $e$ ($s\!<\!0$ on average) it pushes $\hatt$ down (a correction), at mid $e$ it switches off ($s\approx0$), and at high $e$ it reinforces $\hatt>0$. The sign-error rate in the $e<0.1$ region thus drops from $87\%$ (DragonNet) to $63\%$ (DragonNet+Behav), while the true fraction with $\tau>0$ there is only $5\%$.

\begin{figure}[t]
\centering
\begin{minipage}[t]{0.49\textwidth}\centering
\includegraphics[width=\linewidth]{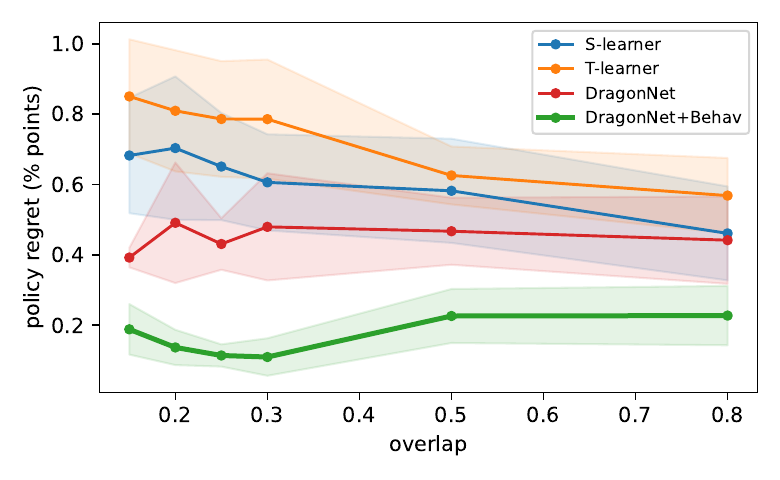}\\[-1pt]
{\footnotesize (a) regret vs \texttt{overlap}}
\end{minipage}\hfill
\begin{minipage}[t]{0.49\textwidth}\centering
\includegraphics[width=\linewidth]{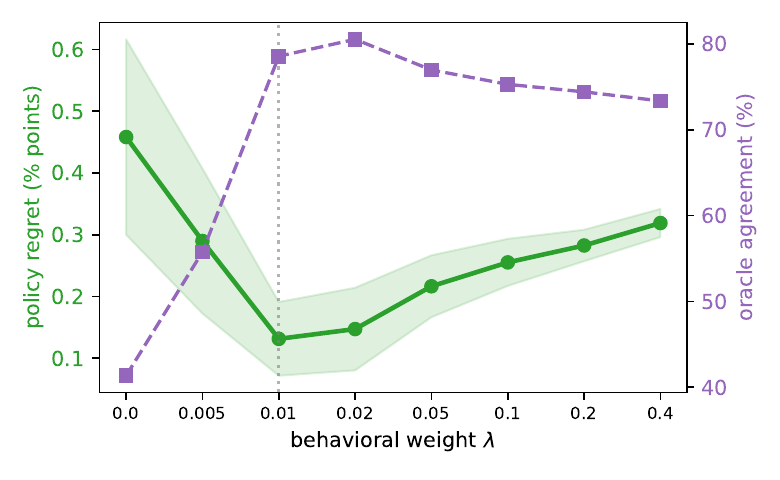}\\[-1pt]
{\footnotesize (b) ablation of $\lambda$ (\texttt{overlap}${=}0.2$)}
\end{minipage}
\caption{Simulation (true CATE known). (a) Policy regret vs overlap (mean~$\pm\,\sigma$): DragonNet+Behav dominates, and the gap grows as overlap drops. (b) Ablation of $\lambda$: U-shaped regret (green) and oracle agreement (purple); an intermediate $\lambda$ minimizes regret, while $\lambda{=}0$ falls back to ``always shoot''.}
\label{fig:sim}
\end{figure}

\paragraph{Results.} Varying the \texttt{overlap} parameter ($12$ repetitions), we compare the S-learner (a single outcome model with T as a feature), T-learner (separate models $\mu_0$ and $\mu_1$ per arm), DragonNet, and DragonNet+Behav. Fig.~\ref{fig:sim}a shows that DragonNet+Behav reaches the lowest \emph{policy regret} $R(\pi)=V(\pi^\star)-V(\pi)$ at every overlap level (e.g.\ $0.19$ pp at \texttt{overlap}${=}0.15$, against $0.39$ for DragonNet and $0.85$ for the T-learner), with a growing advantage as overlap shrinks, confirming that the correction acts exactly where the selection bias is largest. The CATE RMSE is also the lowest, with no extra hyperparameter. This experiment shows that the mechanism is not football-specific: it applies to any regime with an unobserved confounder and low overlap. At fixed overlap (\texttt{overlap}${=}0.2$, Fig.~\ref{fig:sim}b), the regret follows a clear U-shape: $\lambda{=}0$ (DragonNet) shoots in $99\%$ of states (regret $0.46$ pp), an intermediate $\lambda$ ($\approx\!0.01$--$0.02$) minimizes regret ($0.13$ pp, oracle agreement $\approx\!80\%$, shot rate $\approx\!30\%$), and too large a $\lambda$ slowly drifts toward imitating behavior.

\section{Application on real data}
\label{sec:exp}

\paragraph{Data and protocol.} The model is trained and validated (hyperparameters) on several leagues (Jupiler Pro League, Ligue~1, Bundesliga, La Liga, Championship, and Premier League) using synchronized event and tracking data from the 2024--2025 season and the first half of 2025--2026. The data come from Footovision (https://www.footovision.com/). We test on the same leagues, on matches played between the 2025 mid-season break and early April 2026. Each instance is a decision in the finishing zone (a fixed rectangle where about $99\%$ of shots occur), with the graph state $X$ (Sec.~\ref{sec:graph}), the action $T$ (shot vs.\ no shot), and $Y=1$ if a goal is scored within $12$\,s. The holdout set has $n=22{,}507$ decisions, a player shot rate of $31.4\%$, and a goal rate of $6.2\%$. We compare four variants that share the same architecture: TARNet ($\lambda_{\mathrm{prop}}{=}\lambda_{\mathrm{treg}}{=}\lambda{=}0$), DragonNet, Behav only ($\lambda{>}0$ without propensity/targeting), and DragonNet+Behav ($\lambda{=}0.08$, $\beta{=}10$, from cross-validation) ; the holdout set was used only once for final evaluation.

\paragraph{Implementation details.} The shared encoder stacks two GINEConv layers (hidden dimension $64$), a four-part readout, and a three-graph window; the three heads are MLPs ($128$ units), about $80$k parameters in total. Training: RAdam ($\eta{=}10^{-3}$, weight decay $10^{-3}$), batch $512$, up to $70$ epochs with early stopping (patience $15$) on the validation AUC; the train/validation split is done by match to avoid leakage. In the simulation, the encoder is an MLP ($2{\times}64$) and the meta-learners are gradient-boosted forests. The code and evaluation scripts are available online.\footnote{\url{https://github.com/remfootovision/ShouldTheyShoot}}

\paragraph{Metrics.} We separate \emph{factual quality} from \emph{decision quality}. {Factual quality is measured by AUC and Brier score of the outcome and propensity heads. Decision quality is measured by:} (i) the recommended \emph{shot rate} $\PP(\hat\pi{=}1)$, compared to the players' $31.4\%$; (ii) \emph{agreement} with the observed action; (iii) \emph{missed chances} ($\hatt{>}0,T{=}0$) and \emph{ill-advised shots} ($\hatt{<}0,T{=}1$). {We also report the doubly robust (DR) and inverse-propensity-weighting (IPS) off-policy estimators~\cite{dudik2011doubly} as secondary references; both approximate the expected outcome of a policy from observational data, without access to counterfactuals. The DR estimator combines an outcome model with an inverse-propensity correction:}
\begin{equation}
\hat V^{\mathrm{DR}}(\pi)=\tfrac1n\textstyle\sum_i\big[\hmu_{\pi}(X_i)+\tfrac{\1\{T_i=\pi(X_i)\}}{\PP(T_i\mid X_i)}\big(Y_i-\hmu_{T_i}(X_i)\big)\big],
\end{equation}
where $\hmu_\pi(x){=}\pi\hmu_1(x){+}(1{-}\pi)\hmu_0(x)$ and $\PP(T\mid X)$ is estimated by the propensity head $\he$ (clipped to $[0.01,0.99]$); IPS is obtained by dropping the regression term $\hmu$. {However, under low overlap these estimators are unreliable as a \emph{selection} criterion: in low-propensity regions ($e(x)\approx 0$), shots are rare and $\hmu_1$ is inflated by $B(x)$ (Sec.~\ref{sec:sim}), so policies that shoot more receive disproportionately large weights and are over-rewarded. Furthermore, without conditional ignorability, DR and IPS do not identify the causal policy value even asymptotically. We therefore report them for reference only.}

\begin{figure}[t]
\centering
\begin{minipage}[c]{0.45\textwidth}\centering
\includegraphics[width=\linewidth]{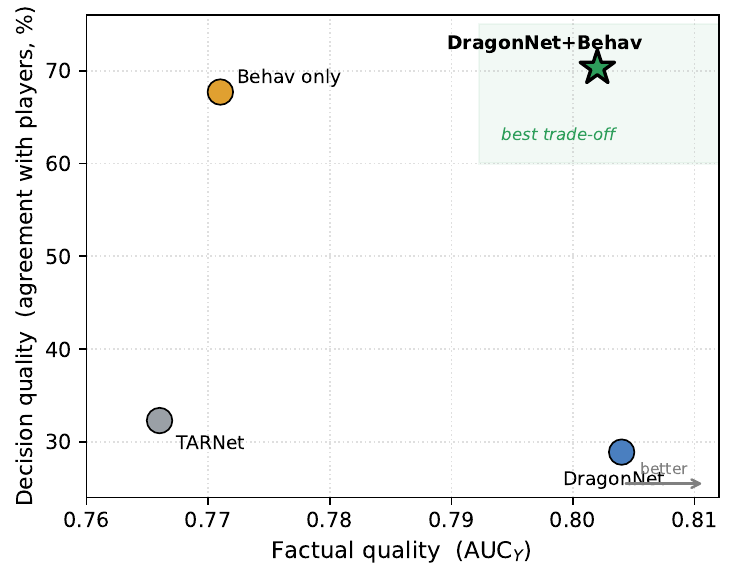}\\[-1pt]
{\footnotesize (a) factual vs decision quality}
\end{minipage}\hfill
\begin{minipage}[c]{0.49\textwidth}\centering
\includegraphics[width=\linewidth]{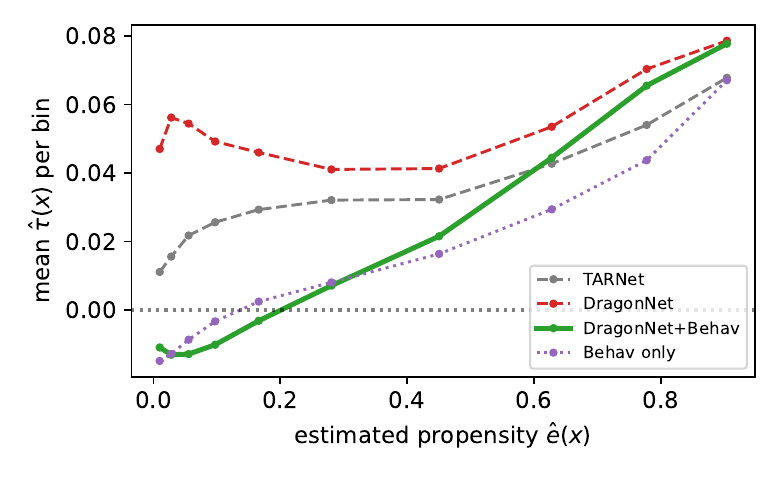}\\[-1pt]
{\footnotesize (b) {observational score vs propensity}}
\end{minipage}
\caption{(a) Factual quality (AUC$_Y$) vs decision quality (agreement with the players) on the holdout, one point per model (numbers from Tab.~\ref{tab:quality}): DragonNet sits bottom-right (good factual, poor decision), Behav only top-left, and only DragonNet+Behav (star) reaches the top-right corner. (b) Mean {estimated observational treatment-effect score} per propensity bin $\he(x)$: behavioral regularization (green) brings $\hatt$ back to a monotone dependence on $\he$ and {reduces the positive scores in} the low-$e$ region (where TARNet/DragonNet {remain} positive).}
\label{fig:tradeoff}
\end{figure}

\paragraph{The factual / decision trade-off.} Tab.~\ref{tab:quality} sums up the central result. DragonNet reaches the best factual quality (AUC$_Y$ $0.804$, Brier $0.0483$) but a pathological policy: it recommends shooting $81.8\%$ of the time, with only $28.9\%$ agreement with the players. Conversely, Behav only gives a reasonable policy (shot $44.4\%$, agreement $67.7\%$) but degrades the factual side (AUC$_Y$ $0.771$, no calibrated propensity head). \textbf{DragonNet+Behav is the only variant, among those we consider, that gets both}: factual quality nearly identical to DragonNet (AUC$_Y$ $0.802$, Brier $0.0488$; $95\%$ CI $[0.790,0.814]$ overlapping DragonNet's $[0.792,0.817]$, a non-significant gap) \emph{and} a realistic policy (shot $39.4\%$, close to the players' $31.4\%$; agreement $70.3\%$). This is exactly the best trade-off: the other models have only one of the two properties. Tab.~\ref{tab:quality} also confirms the large drop in missed chances ($62.5\%\to18.9\%$ between TARNet and DragonNet+Behav) without a blow-up in ill-advised shots, and Fig.~\ref{fig:tradeoff}b shows that behavioral regularization brings the $\hatt$ profile back to a monotone dependence on $\he(x)$. Fig.~\ref{fig:tradeoff}a summarizes the trade-off in a single view: among the variants we consider, DragonNet+Behav is the only one in the top-right corner, combining high factual quality \emph{and} a decision policy close to the players. {We note that the model lift (Tab.~\ref{tab:quality}) is not a reliable criterion here: it is computed by substituting $\hmu$ for unobserved outcomes, precisely in the low-propensity region where $\hmu_1$ is inflated by $B(x)$ (Sec.~\ref{sec:sim}). A policy that shoots more there will appear to gain lift by construction, not because shooting is genuinely better.}

\begin{table}[t]
\centering
\caption{Factual and decision quality on the holdout ($n{=}22{,}507$). Player shot rate: $31.4\%$. Factual metrics $\pm$ bootstrap standard error ($1000$ resamples). Missed chances: $\hatt{>}0$ but $T{=}0$; ill-advised shots: $\hatt{<}0$ but $T{=}1$; lift $=\EE[Y\mid\pi_{\text{mod}}]-\EE[Y\mid\pi_{\text{player}}]$, estimated by substituting $\hmu$ for unobserved outcomes. DragonNet+Behav combines DragonNet's factual quality with a realistic policy.}
\label{tab:quality}
\resizebox{\textwidth}{!}{%
\setlength{\tabcolsep}{3.5pt}
\begin{tabular}{lcccccccc}
\toprule
Model & AUC$_Y\uparrow$ & Brier$_Y\downarrow$ & AUC$_e\uparrow$ & shot\,\% & agree\,\%$\uparrow$ & missed\,\% & ill-adv.\,\% & lift \\
\midrule
TARNet            & $0.766${\scriptsize$\,\pm$.007} & $0.0508${\scriptsize$\,\pm$.0013} & --    & 88.6 & 32.3 & 62.5 & 5.2  & $+0.019$ \\
DragonNet         & $\mathbf{0.804}${\scriptsize$\,\pm$.007} & $\mathbf{0.0483}${\scriptsize$\,\pm$.0012} & 0.902 & 81.8 & 28.9 & 60.7 & 10.4 & $+0.036$ \\
Behav only        & $0.771${\scriptsize$\,\pm$.007} & $0.0501${\scriptsize$\,\pm$.0012} & --    & 44.4 & 67.7 & 22.6 & 9.7  & $+0.008$ \\
\textbf{DragonNet+Behav} & $0.802${\scriptsize$\,\pm$.006} & $0.0488${\scriptsize$\,\pm$.0012} & \textbf{0.904} & \textbf{39.4} & \textbf{70.3} & \textbf{18.9} & 10.8 & $+0.008$ \\
\bottomrule
\end{tabular}}
\end{table}

\paragraph{{Off-policy evaluation under low overlap.}} Tab.~\ref{tab:ope} shows a limitation \emph{of off-policy evaluation itself} under low overlap: DR and IPS favor ``always shoot'' (DR $0.084$) over DragonNet+Behav ($0.076$), because {shots in low-propensity states ($e(x)\approx 0$) receive disproportionately large inverse-propensity weights and $\hmu_1$ is inflated there by $B(x)$, so policies that shoot more are over-rewarded.} {Tab.~\ref{tab:ope} documents this limitation of OPE in our setting, not a ranking of policy quality.} This is one more reason to use a \emph{decision-focused} selection (realistic shot rate, agreement, factual calibration) rather than a biased OPE estimator, in line with work that separates estimation from decision-making~\cite{frauen2025,policy_aligned2025}.

\begin{table}[t]
\centering
\caption{Off-policy evaluation (DR, IPS; $95\%$ bootstrap CIs). Under low overlap, these estimators over-reward ``always shoot'': they should not be the sole selection criterion.}
\label{tab:ope}
\setlength{\tabcolsep}{8pt}
\begin{tabular}{lccc}
\toprule
Policy & shot rate & DR value & IPS value \\
\midrule
DragonNet+Behav & 0.39 & 0.076\,[0.061,\,0.091] & 0.070\,[0.054,\,0.086] \\
DragonNet ($\lambda{=}0$) & 0.82 & 0.084\,[0.069,\,0.099] & 0.074\,[0.061,\,0.089] \\
Always shoot & 1.00 & 0.084\,[0.074,\,0.096] & 0.074\,[0.064,\,0.086] \\
Never shoot   & 0.00 & 0.051\,[0.046,\,0.056] & 0.055\,[0.048,\,0.062] \\
\bottomrule
\end{tabular}
\end{table}

\paragraph{{Behavioral consistency.}} The Spearman correlation between $\he(x)$ and $\hatt(x)$ is $0.70$ ($p\!<\!10^{-3}$): players shoot more where the estimated {observational treatment-effect score} is higher{, confirming that the behavioral relationship generalizes to unseen matches. This is consistent with the assumption underlying~\eqref{eq:behav} and with the description of the problem in Sec.~\ref{sec:problem}: expert players carry weak but useful information about the sign of the treatment effect, and} this positive correlation is exactly the {signal} that behavioral regularization exploits.

\begin{figure}[h!]
\centering
\includegraphics[width=0.82\textwidth]{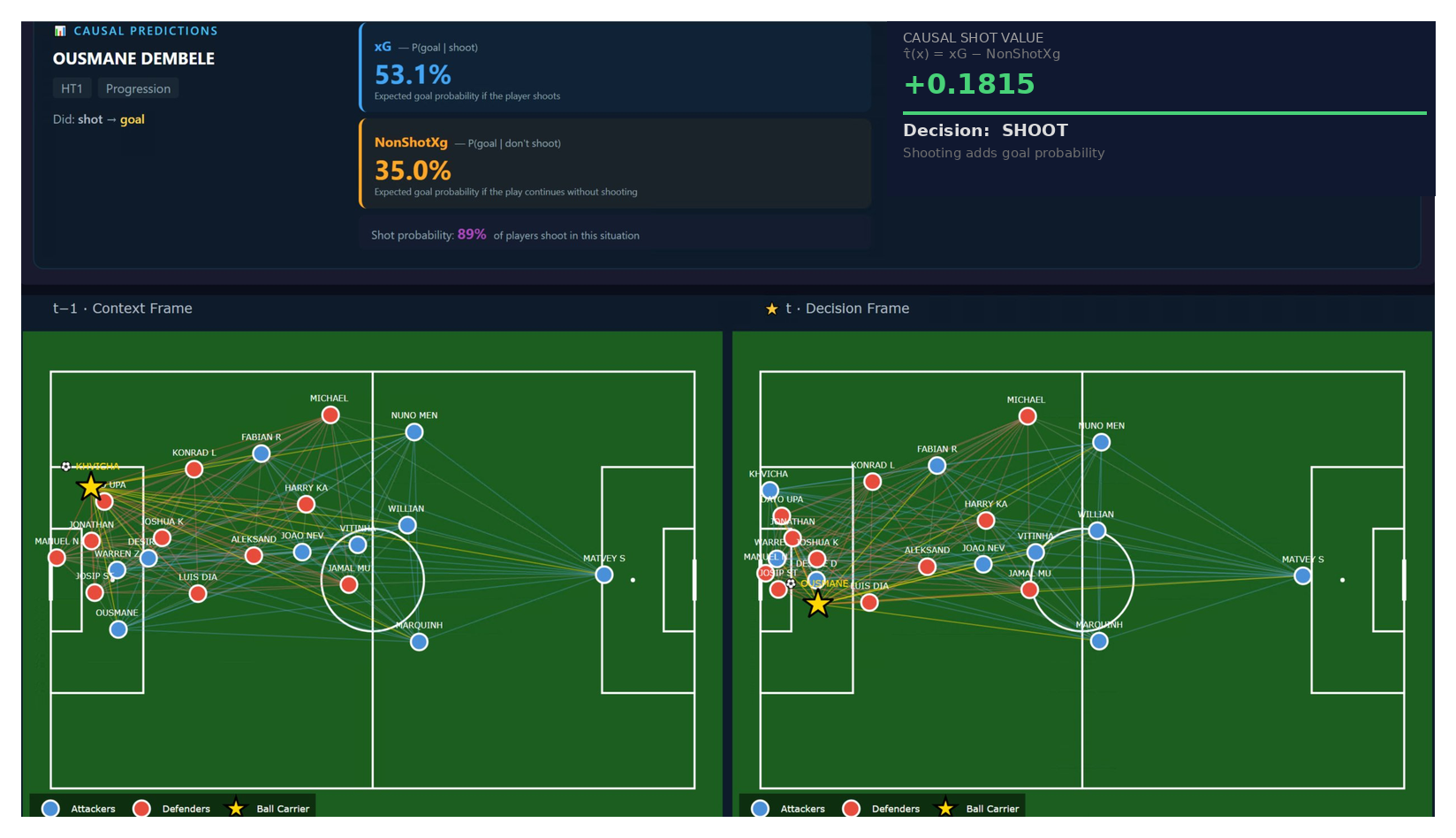}
\caption{Decision-support application. Left: the context frame; right: the decision frame (carrier as a star, attackers/defenders, graph edges). Top: $\hmu_1$ ({outcome score if shot}), $\hmu_0$ ({outcome score if possession continues}), shot probability $\he$, and {decision score} $\hatt=\hmu_1-\hmu_0$ with the recommendation. \textit{Ousmane Dembélé's goal against Bayern Munich in the second leg of the 2026 Champions League semifinal. Recommendation: shoot. xG: 53\%, NonShotXg: 35\%.}}
\label{fig:app}
\end{figure}

\paragraph{Interactive application.} The model is served in an interactive application (Fig.~\ref{fig:app}) that, for each decision, shows {both model-implied outcome scores}, the shot probability, and the {decision score} $\hatt=\hmu_1-\hmu_0$ with a shoot/don't-shoot recommendation. It makes the decision transparent: it separates states where shooting adds goal probability from those where keeping the ball is better, where xG alone stays silent. {Fig.~\ref{fig:app} illustrates the tool on Ousmane Dembélé's chance against Bayern Munich in the 2026 Champions League semifinal: the model estimates $\hmu_1{=}53\%$ and $\hmu_0{=}35\%$ and recommends shooting ($\hatt{=}+0.18$). Because conditional ignorability is not assumed, $\hmu_1$ and $\hmu_0$ are observational model outputs, not identified counterfactual probabilities.}

\section{Discussion and conclusion}
\label{sec:discussion}
We presented a behavioral-regularized causal-inference method for the shooting decision in football, where the observed expert action serves as a directional signal to stabilize the sign of a treatment-effect score under low overlap. The method targets \emph{decision-focused stability}, not the exact calibration of $\tau(x)$: it introduces a deliberate bias (too large a $\lambda$ makes the score behavior-dependent), so model selection must be empirical and policy-focused. Its practical value is that policy learning often fails not because the base estimator is wrong, but because it is unstable where an action is rarely observed; behavioral regularization targets exactly this regime. The output is simple to deploy and easy for an analyst to read, and supports derived player metrics and rankings.

The mechanism is not football-specific. Early prison release is a canonical analogue: judges release sooner the inmates with good behavior, who already have a lower propensity to reoffend, which confounds the inmate's profile with the effect of release. Under low overlap, the few low-release-probability inmates who are nonetheless released form an atypical group whose unobserved traits inflate the apparent effect, exactly the bias $B(x)$ of Sec.~\ref{sec:sim}; behavioral regularization stabilizes the decision rule without claiming to identify the exact causal effect.

The code reproduces Fig.~\ref{fig:sim}; the simulation is fully self-contained, tracking and event data come from Footovision under license, and all nuisances are estimated by cross-validation. {One limitation deserves mention: the behavioral assumption that expert actions carry useful information about the sign of the treatment effect is plausible but untestable from observational data alone; if the sign of $B(x)$ were reversed in some regions, the regularizer would reinforce rather than correct the bias.} Future work includes extending the approach to other decisions (passing, dribbling), off-policy evaluation robust to low overlap, {and empirical tests of the sign of $B(x)$ using data closer to randomization.}

\paragraph*{Acknowledgements} M.O. was partially funded by ANID BECAS/DOCTORADO NACIONAL 21252789

\newpage

\appendix

\section*{Appendix}
\label{app}

\begin{proposition}[Pointwise minimizer]\label{prop:min}
For $x$ with $e(x)\in(0,1)$ and $\beta>0$, the function $v\mapsto\ell_{\mathrm{behav}}(v\mid x)=e(x)\phi_\beta(v)+(1-e(x))\phi_\beta(-v)$ is strictly convex and has the unique minimizer $v^\star(x)=\tfrac{1}{\beta}\mathrm{logit}(e(x))=\tfrac{1}{\beta}\log\tfrac{e(x)}{1-e(x)}$.
\end{proposition}
\begin{proof}
    The second derivative of $\phi_\beta$ is
\begin{equation*}
    \phi_\beta''(u) = \frac{\beta^2 e^{-\beta u}}{(1+e^{-\beta u})^2} > 0 \quad \text{for all } u\in\RR,
\end{equation*}
so $\phi_\beta$ is strictly convex. Since $\ell_{\mathrm{behav}}(\cdot\mid x)$ is a positive linear combination of strictly convex functions, it is strictly convex and therefore has a unique minimizer. Taking the derivative with respect to $v$:
\begin{equation*}
    \frac{\partial}{\partial v}\ell_{\mathrm{behav}}(v\mid x) = -\beta\, e(x)\,\sigma(-\beta v) + \beta\,(1-e(x))\,\sigma(\beta v),
\end{equation*}
where $\sigma(u)=1/(1+e^{-u})$ is the sigmoid. Setting this to zero and using $\sigma(-\beta v)=1-\sigma(\beta v)$:
\begin{align*}
    e(x)\,\sigma(-\beta v) &= (1-e(x))\,\sigma(\beta v)\\
    e(x)\,(1-\sigma(\beta v)) &= (1-e(x))\,\sigma(\beta v)\\
    e(x) &= \sigma(\beta v).
\end{align*}

Applying $\sigma^{-1}$ to both sides:

\begin{equation*}
    v^\star(x) = \frac{1}{\beta}\,\mathrm{logit}(e(x)) = \frac{1}{\beta}\log\frac{e(x)}{1-e(x)}.
\end{equation*}
Since $\ell_{\mathrm{behav}}(\cdot\mid x)$ is strictly convex, this critical point is the unique global minimizer.
\end{proof}
\end{document}